\documentclass[conference]{IEEEtran}

\usepackage{times}
\usepackage{cite}
\usepackage{color}
\usepackage{epsf}
\usepackage{epsfig}
\usepackage{graphicx}
\usepackage{graphics}
\usepackage{epstopdf}
\usepackage[footnotesize]{caption}
\usepackage{amsmath}
\usepackage{amssymb}
\usepackage{amsthm}
\usepackage{amsxtra}
\usepackage[ruled]{algorithm2e}
\usepackage{algorithmic}
\usepackage{enumerate}
\usepackage{multirow}
\usepackage{enumerate}
\usepackage{amssymb}
\usepackage{lipsum}
\usepackage{setspace}
\usepackage{multirow}
\usepackage{mathcomp}
\usepackage{epstopdf}

\usepackage{ifpdf}
\ifpdf
  \usepackage{epstopdf}
\fi

\pdfoutput=1

\begin{document}
\title{Exploiting QoS Flexibility for Smart Grid and IoT Applications Using TV White Spaces: \\ Extended Version}
\author{Naveed UL Hassan, Chau Yuen and Muhammad Bershgal Atique
\thanks{Naveed UL Hassan and Muhammad Bershgal Atique are with the Electrical Engineering Department at the Lahore
University of Management Sciences (LUMS), Lahore, Pakistan 54792. (Email: naveed.hassan@yahoo.com, 17100149@lums.edu.pk).} 
\thanks{Chau Yuen is with the Engineering Product Development at the Singapore University of Technology and Design (SUTD), 8 Somapah Road, Singapore 487372. (Email: yuenchau@sutd.edu.sg).} 
\thanks{This research is partly supported by Lahore University of Management Sciences (LUMS) and A*Star Singapore SERC project number 142 02 00043.}
}
\IEEEoverridecommandlockouts
\maketitle
\begin{abstract}
In this paper, we consider the utilization of TV White Spaces (TVWS) by small Cognitive Radio (CR) wireless network operators (SCWNO) to support the communication needs of various smart grid and internet of things (IoT) applications. Spectrum leasing could be essential for SCWNO to support a wide range of applications. In our paper, we assume that in order to ensure Quality of service (QoS) requirements, the CR operator has the option of leasing the specially designated high priority TVWS channels (HPC) for short term use by paying a small fee according to the actual usage. Based on Lyapunov drift plus penalty function framework, we develop an online algorithm to exploit QoS flexibility. Such flexibility might be available in terms of data transmission delays as well as data quality reduction in order to minimize the overall HPC leasing cost. The developed algorithm also provides three adjustable parameters that could be controlled to tradeoff the total cost and QoS. The performance of the proposed algorithm is very close to the optimal offline lower bound solution.
\end{abstract}

\section{Introduction}\label{sec:int}
The significance of Cognitive Radio (CR) technology that allows the opportunistic and intelligent use of wireless spectrum is well recognized \cite{cr_1,cr_2,cr_3,cr_4}. In recent years, TV White spaces (TVWS) have also become available for wireless communication due to the transition from analog to digital TV broadcasts. TVWS comprise of relatively large frequency blocks in the very high frequency (VHF) and ultra high frequency (UHF) bands with a tremendous potential to support several new applications and services. There is a general consensus among the spectrum regulators to provide an unlicensed access to TVWS~\cite{fcc,of_com,sing_ida}. There are also some efforts to standardize the use of TVWS spectrum. For example, the IEEE 802.11af, also sometimes referred to as `White-Fi' or `Super Wi-Fi' adds the TVWS bands to the IEEE 802.11 family of specifications.

Fortunately, the free TVWS availability has coincided with the ever increasing interest in smart grids, internet of things (IoT), and machine-to-machine (M2M) communication \cite{new_m2m1,new_m2m2}. The data rate requirements of many novel smart grid applications, e.g., demand response management, energy monitoring for shared facilities (e.g. lift, corridor light), environmental sensing, etc., could be supported by exploiting TVWS. Some of these applications require stringent Quality of Service (QoS) guarantees. Using the CR technology and TVWS, many new small and virtual wireless data service providers can also emerge to support these applications. We would refer to these operators as `small cognitive radio wireless network operators' (SCWNO). The competition would also increase among the traditional wireless operators and SCWNOs for the freely available TVWS. Hence, to support a wide range of applications with varying QoS requirements, there are also some proposals to assign certain channels in the TVWS spectrum as high priority channels (HPCs)~\cite{sing_ida}, which could be temporarily leased by the interested CR operators by paying a small license fee. In this context, minimizing the HPC leasing cost while providing the desired QoS guarantees would become a challenge, particularly for the SCWNOs. 
 
SCWNO can minimize HPC leasing costs by exploiting the maximum amount of flexibility that could be offered by the considered smart grid or IoT application \cite{mag_new}. For example, for certain applications, SCWNO can wait for free TVWS spectrum availability and delay the transmission of data packets according to the prescribed requirements. Simultaneously, SCWNO can further exploit the redundancy in the measurements (e.g., smart meter data readings, environmental sensor readings, etc.) and it can reduce the size of data packets (e.g., by dropping the least significant bits). The transmission of some reduced size packets in different time slots would impact the temporal quality of data but it would also reduce the HPC leasing cost. Therefore, SCWNO is required to intelligently exploit all the available QoS flexibility (delay and quality) to minimize its spectrum leasing costs. 



In \cite{icc_16}, we considered this problem and we developed an optimal offline algorithm with the objective of minimizing the overall HPC leasing cost of the SCWNO. We imposed hard constraints on the transmission delays and data quality. We also developed an online heuristic whose performance was not close enough to the optimal offline algorithm. In current work, we use Lyapunov drift plus penalty framework, which has been used by several researchers to obtain real time solutions of many interesting research problems \cite{lup_1,lup_2,lup_3,lup_4}. For our problem, using this framework, we decouple the problem in time and develop an efficient and novel online algorithm that can be independently solved in each time slot. The performance of our Lyapunov based online algorithm is very close to the optimal lower bound that is provided by the offline algorithm in \cite{icc_16}. The developed algorithm also provides three controllable parameters that could be tuned to tradeoff HPC leasing cost, delay and data quality according to the application requirements. In practical systems, this algorithm can more efficiently exploit the QoS flexibility and further reduce the HPC leasing cost of the SCWNOs.

The rest of the paper is organized as follows. In Section~\ref{sec:sys} we discuss the system model and formulate the problem, in Section~\ref{sec:alg} we develop our online algorithm, in Section~\ref{sec:sim} we present the simulation results, while the paper is concluded in Section~\ref{sec:conc}.

\section{System Model and problem formulation}
\label{sec:sys}
\subsection{System Model}
Consider the scenario as shown in Figure 1. The data from several smart homes in a residential community and multiple buildings in a smart city is aggregated (e.g., using power line communication) at the local data concentrators deployed by the SCWNO. Using the CR technology, the data concentrators transmit the queued data on TVWS to the remote SCWNO Base Station (BS). From the remote BS, the data is further transmitted on the backhaul internet for smart grid and IoT services. 
\begin{figure}
\centering
\includegraphics[width=0.48\textwidth,height=.25\textheight]{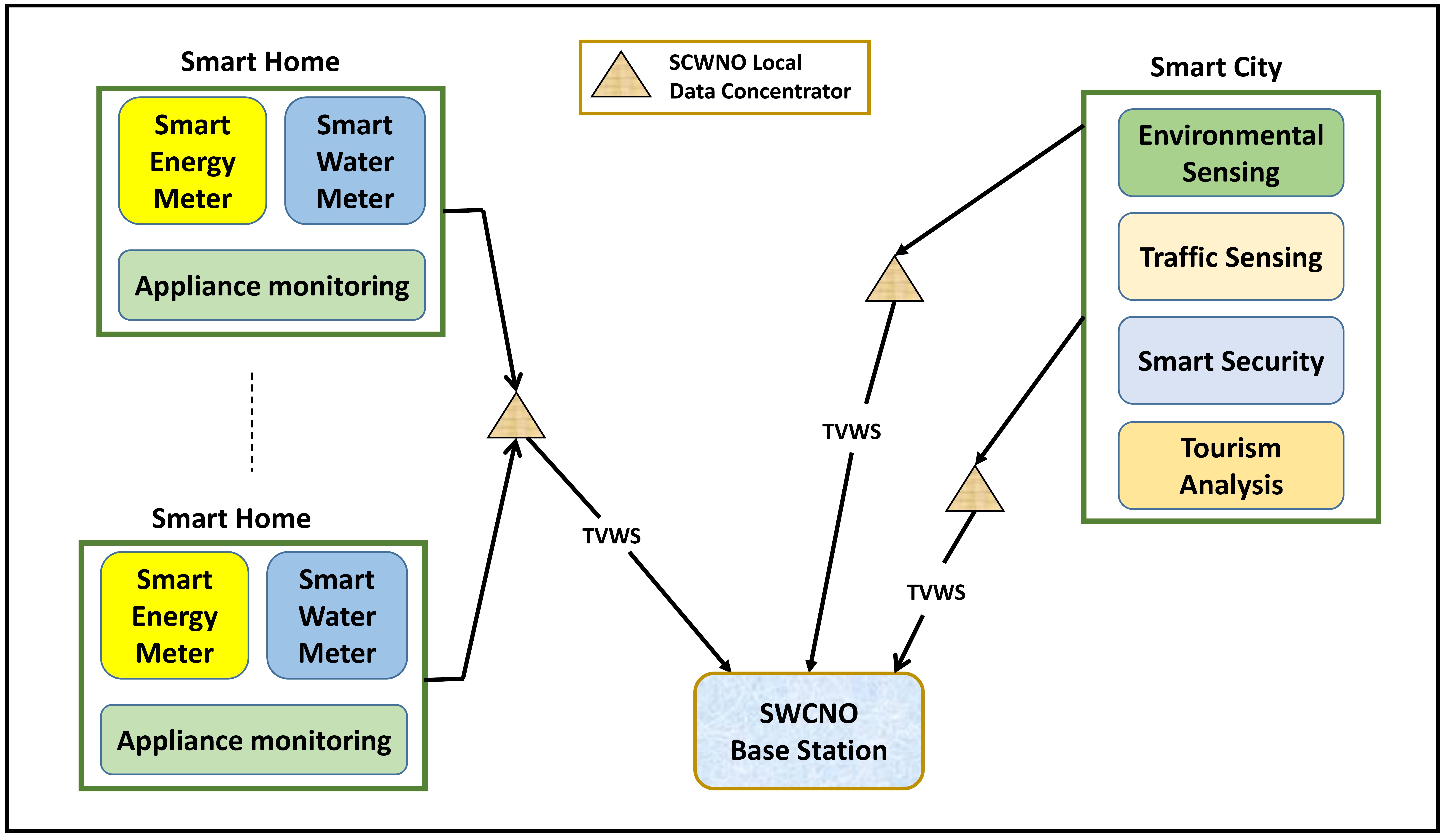}
\caption{SCWNO exploiting TVWS for smart grid \& IoT applications}
\label{fig:fig1}
\end{figure} 

For such applications, we assume a discrete time system and we consider the transmission of data queued at the local data concentrators using TVWS. In the remainder of this work, our focus will be on any one such data concentrator. We assume that `data units' arrive or depart from the data concentrator queue. Each data unit comprises of multiple packets and each packet comprises of multiple bits. The queue length, denoted by $Q(t)$, evolves according to the following update equation, 
\begin{equation}
Q(t+1)=\max(Q(t)-R(t),0)+A(t+1)
\label{queue:eqn}
\end{equation}
In this equation, $A(t)$ denotes the arrival rate in time slot $t$, while the departure rate from the queue at time $t$ is denoted by $R(t)$. The arrival and departure rates are defined in terms of data units. We assume that only one data unit can arrive in the queue at any time $t$, i.e., $A(t) \in \{0,1\}, \forall t$, such that $A_{\text{max}}=1$. Similarly, we also assume that a maximum of only one data unit (either of full size or reduced size) can be transmitted in any given time slot $t$, i.e., $R_{\text{max}}=1$ and subsequently, $R(t) \in \{0,1\},\: \forall t$. 

Let, $M$ denote the total number of data bits in one full size data unit. We suppose that the number of bits in a reduced size data unit are $\alpha M$, where $0 < \alpha < 1$. A reduced size data unit might be obtained by dropping $(1-\alpha) M$ least significant bits. Let, $c(t)$ denotes the time varying per unit HPC spectrum price to transmit one data bit in time slot $t$. We assume that the spectrum price increases linearly with the number of bits. Therefore, the size of data unit also becomes an important factor in determining the cost of purchased spectrum. The reduction in the size of data unit has an adverse impact on the overall data quality. However, it is important to note that for some applications in the scenario considered in Figure \ref{fig:fig1}, e.g., smart energy and water meters there is a certain amount of redundancy in the consecutive readings. This redundancy could allow transmitting reduced size data units in some time slots to avoid HPC purchase decisions. Thus, in any time slot $t$, the SCWNO BS has the options to make any one of the following decisions:
\begin{enumerate}	
	\item Transmission of one full size data unit using the freely available TVWS spectrum.
	\item Transmission of one reduced size data unit using the freely available TVWS spectrum.
	\item Transmission of one full size data unit by purchasing HPCs.
	\item Transmission of one reduced size data unit by purchasing HPCs.
	\item No transmission of a data unit.
\end{enumerate}

\subsection{Problem Formulation}
Let $d_f^f(t)$, $d_f^r(t)$, $d_p^f(t)$, and $d_p^r(t)$ respectively denote the binary decision variables for the transmission/non-transmission of one data unit of full size or reduced size on free TVWS and HPC spectrum. For example, $d_f^f(t)=1$ indicates the transmission of one full size data unit using the freely available TVWS spectrum at time $t$. These decision variables are mutually exclusive, i.e., in any time slot $t$, only one decision variable can become 1. Additionally, setting all the decision variables equal to 0 would indicate non-transmission of a data unit in time slot $t$. In terms of the decision variables and the fact that only one data unit may be transmitted in any given time slot, the departure rate $R(t)$ can be written as, 
\begin{equation}
R(t)=d_f^f(t) + d_f^r(t) + d_p^f(t) + d_p^r(t), \quad \forall t
\label{dep:rate}
\end{equation}

The transmission of a full size or a reduced size data unit on freely available TVWS spectrum does not cost the SCWNO. However, if there is a strong competition for TVWS spectrum among various operators, free TVWS availability will become time varying and limited. In such scenarios, in order to maintain a certain QoS, in terms of both delay and data quality, the SCWNO might be required to lease HPCs. Our objective in this paper is to minimize the time average expected cost incurred due to the HPC purchase decisions. We can mathematically write down the objective function as:
\begin{equation}
\mathcal{C}= \lim_{t \rightarrow \infty} \frac{1}{t} \sum_{\tau=0}^{t-1} \mathbb{E}\left[\left(d_p^f(\tau) M + d_p^r(\tau) \alpha M\right) c(\tau) \right]
\label{cost:fun0}
\end{equation}
To further simplify our objective function, let $c^f(t)=M c(t)$ and $c^r(t)=\alpha M c(t)$. We can now re-write \eqref{cost:fun0} as,
\begin{equation}
\mathcal{C}= \lim_{t \rightarrow \infty} \frac{1}{t} \sum_{\tau=0}^{t-1} \mathbb{E} \left[d_p^f(\tau) c^f(\tau) + d_p^r(\tau) c^r(\tau) \right]
\label{cost:fun}
\end{equation}
We have the following optimization problem,
\begin{equation}
\min_{d_f^f(t), d_f^r(t), d_p^f(t), d_p^r(t)} \: \mathcal{C}
\label{obj:1}
\end{equation}
subject to the following constraints:
\begin{equation}
d_f^f(t) + d_f^r(t) + d_p^f(t) + d_p^r(t) \leq 1, \quad \forall t
\label{mut:excl}
\end{equation}
\begin{equation}
d_f^f(t) \in \{0,1\}, \: d_f^r(t) \in \{0,1\}, \: d_p^f(t) \in \{0,1\}, \: d_p^r(t) \: \in \{0,1\}, \: \forall t
\label{bin:const}
\end{equation}
\begin{equation}
\lim \sup_{t \rightarrow \infty} \frac{1}{t} \sum_{\tau=0}^{t-1} \mathbb{E} \left[Q(\tau) \right] < \infty
\label{ave:Q}
\end{equation}
Constraint \eqref{mut:excl} indicates that in any time slot $t$, a maximum of only one decision variable can become 1. Constraint \eqref{bin:const} is due to the binary nature of decision variables. For stability of the queue, the time average queue backlog should be finite, which is given as constraint \eqref{ave:Q}. This problem imposes certain conditions on the arrival and departure dynamics and on queue stability. However, there are no constraints on the delay and data quality. 

In the following, we will first define two virtual queues, respectively known as the quality-aware and delay-aware queues. Quality-aware queue keeps track of the data quality, while delay-aware queue keeps track of queuing delays. We will then use Lyapunov drift plus penalty function framework to obtain a real time algorithm to minimize the cost function.

\section{Online Algorithm Development}\label{sec:alg}
\subsection{Lyapunov drift plus penalty framework}
We have two QoS dimensions, i.e., data quality (measured in terms of the number of data units whose size is reduced) and queuing delay.
We define two virtual queues, which are denoted by $Y(t)$ and $Z(t)$. Virtual queue, $Y(t)$ is the quality-aware queue, while $Z(t)$ is the delay-aware queue. 
\begin{equation}
Y(t+1)=\max(Y(t)-R(t) + \epsilon_q \left[d_f^r(t)+d_p^r(t)\right] ,0)
\label{qual:q}
\end{equation}
\begin{equation}
\begin{split}
Z(t+1)= & \max(Z(t)-R(t) + \\ & \epsilon_d \left[1-\left(d_f^f(t) + d_f^r(t) + d_p^f(t) + d_p^r(t)\right) \right],0)
\end{split}
\label{qual:d}
\end{equation}
Both the virtual queues have the same departure rate $R(t)$ as the original queue $Q(t)$. However, whenever we transmit a reduced size data unit, which reduces data quality, $Y(t)$ is incremented by $\epsilon_q>0$. Similarly, whenever there are no transmissions from the non-empty queue, which increases the queuing delay, $Z(t)$ is incremented by $\epsilon_d >0$. 

Let, $\Theta(t)=\left(Q(t),Y(t),Z(t)\right)$ denote the concatenated vector of actual and virtual queues. We define the following Lyapunov function,
\begin{equation}
\mathcal{L}(\Theta(t))=\frac{1}{2} \left[Q^2(t)+Y^2(t)+Z^2(t) \right]
\label{lyap:1}
\end{equation}
The Lyapunov function is a measure of congestion in the actual and virtual queues. We define the conditional 1-slot Lyapunov drift function as,
\begin{equation}
\Delta (\Theta(t))=\mathbb{E} \left[\mathcal{L}(\Theta(t+1))-\mathcal{L}(\Theta(t)) | \Theta(t) \right]
\label{cond:lyap}
\end{equation}
The Lyapunov drift plus penalty function is given by the following expression,
\begin{equation}
\Delta (\Theta(t))+V \mathbb{E} \left[d_p^f(t) c^f(t) + d_p^r(t) c^r(t) | \Theta(t) \right]
\label{bound:express}
\end{equation} 
where, $V>0$ is a parameter that can be used in the delay, quality and performance tradeoffs.
Using the Lyapunov drift plus penalty function framework, instead of minimizing the original objective function, we can minimize an upper bound on \eqref{bound:express}, which is given by the following function, 
\begin{equation}
\begin{split}
\mathcal{G}(t)= & V \left\{d_p^f(t) c^f(t)+ d_p^r(t) c^r(t) \right\}  \\ &
+ Y(t) (\epsilon_q-1) \left\{d_f^r(t)+d_p^r(t) \right\}  \\ &
- Y(t) \left\{d_f^f(t)+d_p^f(t) \right\}  \\ &
- Q(t) \left\{d_f^f(t) + d_f^r(t) + d_p^f(t) + d_p^r(t) \right\}  \\ &
- Z(t) (\epsilon_d+1) \left\{d_f^f(t) + d_f^r(t) + d_p^f(t) + d_p^r(t) \right\}
\end{split}
\label{master:eqn}
\end{equation}
The detailed derivation of \eqref{master:eqn} is provided in Appendix A. Now, instead of solving the original optimization problem, the decoupling in time provided by \eqref{master:eqn} can be used to develop an online algorithm. Therefore, in every time slot $t$, we have the following optimization problem: 
\begin{equation}
\min_{d_f^f(t), d_f^r(t), d_p^f(t), d_p^r(t)} \ \mathcal{G}(t)
\label{fin:opt}
\end{equation}
subject to constraints \eqref{mut:excl}, \eqref{bin:const}. In the next subsection, we develop an online algorithm to obtain an optimal solution of problem \eqref{fin:opt}.

\subsection{Online Algorithm}
Let, $h(t) \in \{0,1,2\}$, indicate the availability/non-availability of free TVWS spectrum in time slot $t$. The value of $h(t)=0$ indicates the non-availability of free TVWS spectrum for the transmission of either a full size or a reduced size data unit. The value of $h(t)=1$, on the other hand, indicates the availability of free TVWS spectrum, which is only sufficient for the transmission of one reduced size data unit. Finally, the value of $h(t)=2$, indicates the availability of free TVWS spectrum, which is enough for the transmission of one full size data unit. Depending on the availability of free TVWS spectrum, HPC prices and actual and virtual queue lengths, we can develop an algorithm that solves the optimization problem \eqref{fin:opt} and makes appropriate real time decisions on the purchase of HPCs and the reduction in the size of data units. 

In our optimization problem, constraint \eqref{mut:excl}, \eqref{bin:const} ensures that no more than one binary decision variables can be 1. By exploiting these two constraints, we obtain four possible values of the objective function \eqref{fin:opt}. These values can be obtained by setting only one binary decision variable equal to 1,
\begin{equation}
g_1(t)=-\left\{Q(t)+Z(t)(\epsilon_d+1)+Y(t)\right\}
\label{val:1}
\end{equation}
\begin{equation}
g_2(t)=\epsilon_q Y(t)-\left\{Q(t)+Z(t)(\epsilon_d+1)+Y(t)\right\}
\label{val:2}
\end{equation}
\begin{equation}
g_3(t)=V c^f(t)-\left\{Q(t)+Z(t)(\epsilon_d+1)+Y(t)\right\}
\label{val:3}
\end{equation}
\begin{equation}
g_4(t)=\left\{V c^r(t)+\epsilon_q Y(t)\right\}-\left\{Q(t)+Z(t)(\epsilon_d+1)+Y(t)\right\}
\label{val:4}
\end{equation}
Here, $g_1(t)$ is obtained by letting $d_f^f(t)=1$ in \eqref{fin:opt}. Similarly, $g_2(t)$, $g_3(t)$, and $g_4(t)$ are respectively obtained by setting $d_f^r(t)$, $d_p^f(t)$ and $d_p^r(t)$ equal to 1 in \eqref{fin:opt}. It is obvious that the value of $g_1(t) \leq 0$, while the remaining values could be positive, negative or zero depending on the dynamic HPC prices, queue lengths and different parameters. We also define a variable denoted by $g_5(t)$ and set it equal to 0 to identify the case when the transmitter decides not to transmit a data unit of any size in time slot $t$.

There are three possible states of free TVWS spectrum availability. When $h(t)=2$, we have sufficient freely available TVWS spectrum to transmit a full size data unit. In this case, if the queue length is non-zero, the value of $g_1(t)$ will always be negative and minimum. Therefore, the obvious decision would be to exploit the availability of free TVWS spectrum to transmit one full size data unit. When $h(t)=1$, we have sufficient free TVWS spectrum to transmit a reduced size data unit. However, our decision to use the free TVWS spectrum for the transmission of one reduced size data unit also depends on the quality and delay. In this case, we can also note that $g_4(t) > g_2(t)$ by a factor of $V c^r(t)$, which indicates that if we decide to transmit a reduced size data unit then we should always transmit it on the free TVWS spectrum. Thus, if the queue length is non-zero, we can either exploit the free TVWS spectrum and transmit a reduced size data unit or we can purchase HPC spectrum to transmit one full size data unit or we do not transmit at all. The final decision among these three options will be made depending on the actual values of $g_i(t), i=2,3,5$ in time slot $t$. We will select the option that gives the minimum value. When $h(t)=0$, we do not have freely available TVWS spectrum. In this case, there are three possible options, purchase HPC spectrum for the transmission of a full size data unit, or purchase HPC spectrum for the transmission of a reduced size data unit or no transmission of data unit in time slot $t$. Now, we have to decide based on the values of $g_i(t), i=3,4,5$, and select the option that gives the minimum value. All these steps are detailed in Algorithm \ref{online:cost}.    

\begin{algorithm}[htb]
\caption{Lyapunov Framework Based Online Algorithm}
\label{online:cost}
Initialize variables and set the values of parameters $V>0$, $\epsilon_q >0$, $\epsilon_d >0$ and $0 < \alpha < 1$, depending on the QoS requirements of the application.
\begin{algorithmic}[1]
\STATE At the start of every time slot $t$, observe the values of $c^f(t)$, $c^r(t)$, $Q(t)$, $Y(t)$, $Z(t)$ and $h(t)$.
\STATE Initialize: $d_f^f(t)=0$, $d_f^r(t)=0$, $d_p^f(t)=0$ and $d_p^r(t)=0$.
\STATE Compute the values of $g_i(t), i=1,2,3,4$ using \eqref{val:1}, \eqref{val:2}, \eqref{val:3} and \eqref{val:4} and set $g_5(t)=0$.
\IF{$h(t)=2$}
\STATE Set $d_f^f(t)=1$, if $Q(t) > 0$. 
\ENDIF
\IF{$h(t)=1$}
\STATE Find $g*=\min \:[g_2(t), g_3(t), g_5(t)]$. 
\STATE Set $d_f^r(t)=1$, if $g_2(t)=g*$ and $Q(t) >0$.
\STATE Set $d_p^f(t)=1$, if $g_3(t)=g*$ and $Q(t) >0$.
\ENDIF
\IF{$h(t)=0$}
\STATE Find $g*=\min \:[g_3(t), g_4(t), g_5(t)]$. 
\STATE Set $d_p^f(t)=1$, if $g_3(t)=g*$ and $Q(t) >0$.
\STATE Set $d_p^r(t)=1$, if $g_4(t)=g*$ and $Q(t) >0$.
\ENDIF
\STATE Update the values of $Q(t)$, $Y(t)$ and $Z(t)$ using \eqref{queue:eqn}, \eqref{qual:q} and \eqref{qual:d} for the next time slot.
\STATE Repeat by going to step 1.
\end{algorithmic}
\end{algorithm}

This algorithm also provides various controllable parameters, i.e., $V$, $\epsilon_d$ and $\epsilon_q$ that could be used to tradeoff cost, delay and data quality.

\subsection{Analytical Bounds on worst case delay and data quality}\label{sec:anal}
We have one actual queue $Q(t)$ and two virtual queues, i.e., $Y(t)$ and $Z(t)$. Let $c^f_{\text{max}}$ and $c^r_{\text{max}}$ respectively denote the maximum price to purchase full size and the reduced size data units. Let, $Q_{\text{max}}$, $Y_{\text{max}}$ and $Z_{\text{max}}$ respectively denote the upper bounds on $Q(t)$, $Z(t)$ and $Y(t)$. These bounds are given by the following expressions,
\begin{equation}
Q_{\text{max}} = V c^f_{\text{max}} + 2
\label{act:Q_b}
\end{equation}
\begin{equation}
Z_{\text{max}} = \frac{V c^f_{\text{max}}}{1+\epsilon_d} + \epsilon_d
\label{act:Z_b}
\end{equation}
\begin{equation}
Y_{\text{max}} = V c^f_{\text{max}} + \epsilon_q
\label{act:Y_b}
\end{equation}
Let, $D_{\text{max}}$ denote the worst case delay bound. Since, we assume that only one data unit can enter or leave the queue in any time slot, the worst case delay can be obtained from the maximum queue length, i.e.,
\begin{equation}
D_{\text{max}}=\left\lceil V c^f_{\text{max}} +1 \right\rceil
\label{del:bound}
\end{equation}
where, $\left\lceil . \right\rceil$ denotes the ceil operation. 
A data unit arriving in the queue at time $t$ will always leave the queue before $t+D_{\text{max}}+1$ time slot.

Finally, we can upper bound the maximum number of reduced size data units in every $D_{\text{max}}$ time slots by the following equation,
\begin{equation}
S_{\text{max}} = \left\lceil \min \left(\left(\frac{Y_{\text{max}}+D_{\text{max}}}{\epsilon_q} \right),D_{\text{max}}\right) \right\rceil
\label{qualy:bound}
\end{equation} 
For example, if $D_{\text{max}}=100$, then the total number of reduced size data units in every 100 data units are given by \eqref{qualy:bound}. If $D$ denotes the total number of time slots of operation (e.g., $D=10,000$ time slots), then an upper bound on the total number of reduced size data units out of total $D$ data units (recall we have one data unit arriving every time slot) is determined by the following expression,
\begin{equation}
S_{\text{max}}^D = \min \left(\left\lceil \frac{D S_{\text{max}}}{D_{\text{max}}}  \right\rceil, D \right)
\label{tot_q:bound}
\end{equation}
The proofs of \eqref{act:Q_b}, \eqref{act:Y_b}, \eqref{act:Z_b}, \eqref{del:bound} and \eqref{qualy:bound} are given in Appendix B.

\section{Simulation Results}\label{sec:sim}
The availability of free TVWS spectrum is generated in MATLAB as uniformly distributed integer random variables. The random variable takes three possible values to indicate the non-availability of free TVWS spectrum, the availability of free TVWS spectrum for the transmission of one reduced size data unit and the availability of free TVWS spectrum for the transmission of one full size data unit. We assume $\alpha=0.5$, i.e., the size of a reduced size data unit is half that of the full size data unit. The values of $c^f(t)$, i.e., the cost of leasing HPC for the transmission of one data unit is generated as a uniformly distributed random variable between $[0.5,5] \$\text{cents}$. The values of $c^r(t)=0.5 \times c^f(t)$. In these simulations, we compare the performance of our proposed online algorithm with the optimal offline bound and the online heuristic that were developed in \cite{icc_16}. For fair comparison with the schemes presented in \cite{icc_16} we assume $A(t)=1, \forall t$, i.e., a data unit arriving in every time slot.   

\begin{figure}[htb]
\centering
\includegraphics[width=0.48\textwidth,height=.25\textheight]{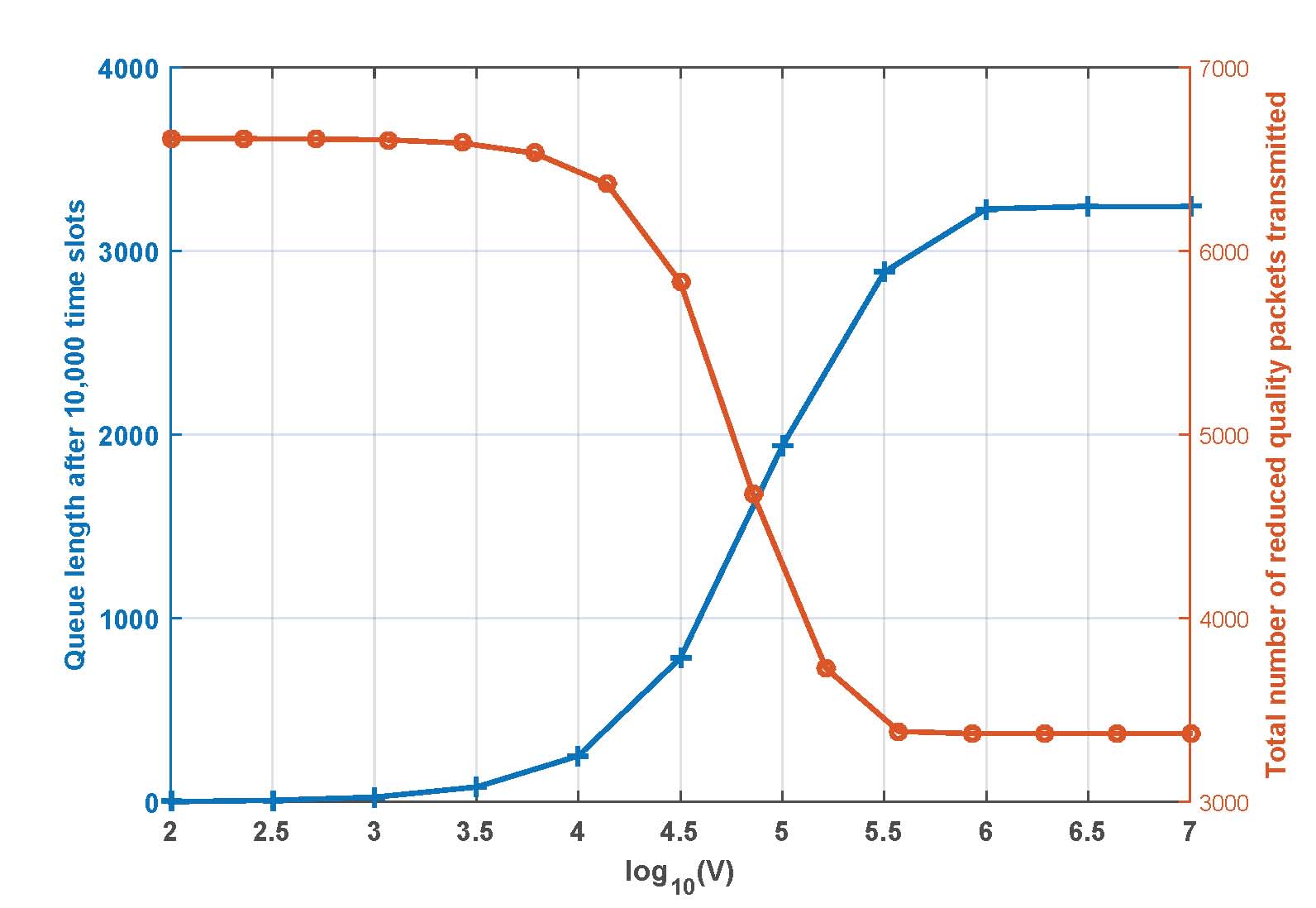}
\caption{Queue length and the number of reduced packets for different values of $V$ (due to large values of $V$, we have plotted $\log_{10}(V)$}
\label{fig:figa}
\end{figure}

In Figure \ref{fig:figa}, we fix $\epsilon_q=\epsilon_d=1$ and we vary the value of parameter $V$. We run the simulation for $D=10,000$ time slots. We plot the resulting queue length at the end of 10,000 time slots and on the same graph, we also plot the total number of data units that are transmitted with reduced size. As we consider a large range of parameter $V$, the horizontal axis is in $\log_{10}$ scale. Let $S$ denote the total number of data units that are transmitted with reduced size. Similarly, let the total number of transmitted data units be denoted by $N=D-Q(D)-1$. We can equivalently interpret the results in Figure \ref{fig:figa} as the transmission of $N$ data units out of which the size of $S$ data units can be reduced and these $N$ data units have to be transmitted in $D$ time slots. We can see that for small values of $V$ the queue length is small but as the value of $V$ increases the queue length also increases. Similarly, as the value of $V$ increases, the total number of packets that are transmitted with reduced size decreases. 

\begin{figure}[htb]
\centering
\includegraphics[width=0.48\textwidth,height=.25\textheight]{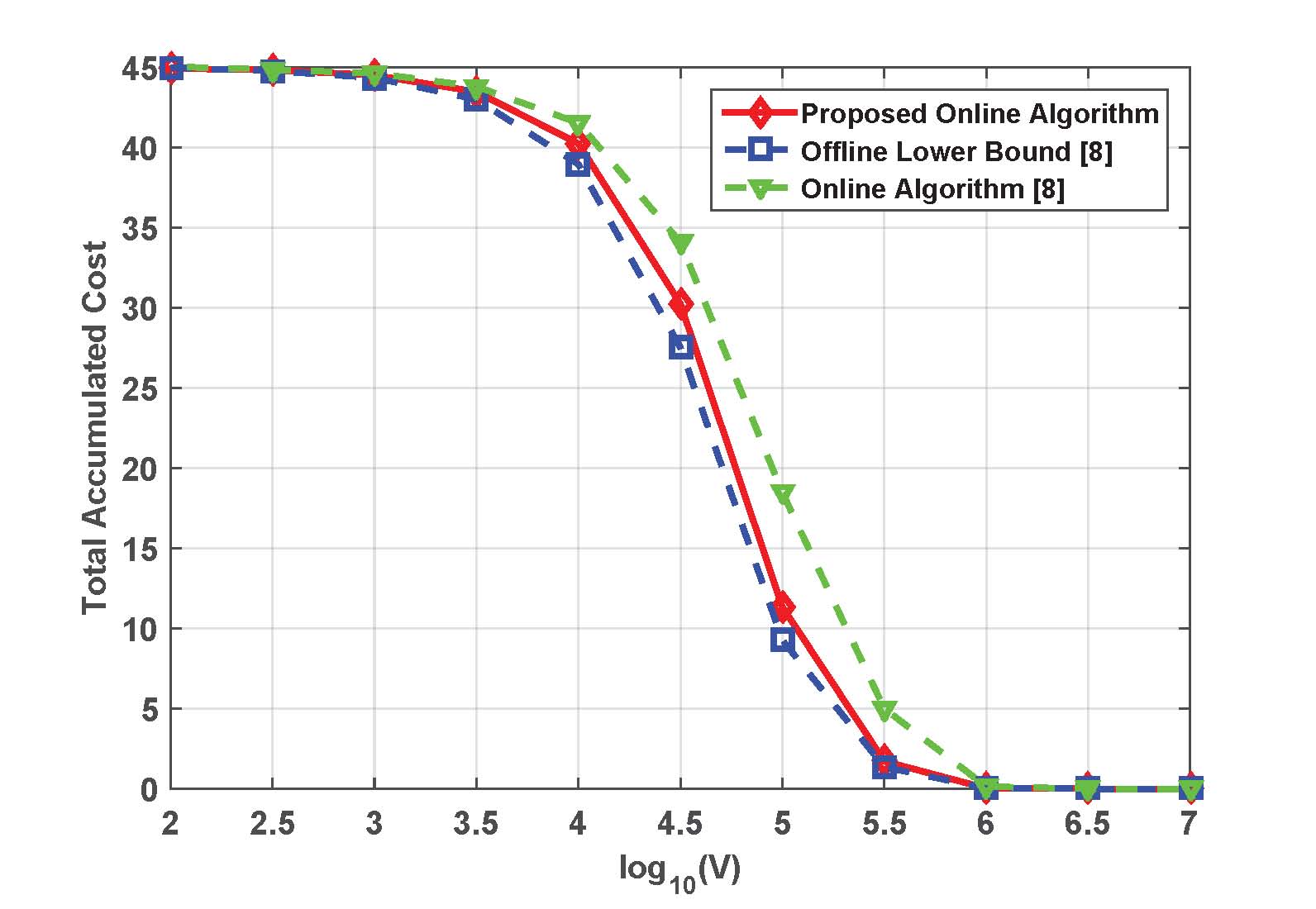}
\caption{Total accumulated cost for different values of parameter V (due to large values of $V$, we have plotted $\log_{10}(V)$}
\label{fig:figb}
\end{figure}

In Figure \ref{fig:figb}, we plot the total cost of our proposed algorithm, the optimal offline lower bound and the online heuristic in \cite{icc_16} for different values of parameter $V$. For every value of $V$, for comparison with the algorithms in \cite{icc_16} we have used $S$ and $N$ that were obtained in Figure \ref{fig:figa}. We can observe that when the value of $V$ is small, the total cost is very high. Increasing the value of $V$ reduces the total cost, and it goes to zero for very large values of $V$. However, at large values of $V$, the corresponding values of $S$ and $Q(D)$ are also high. The performance of our proposed Lyapunov algorithm is very close to the optimal lower bound for all the values of $V$. The mean square difference between the total cost of our proposed algorithm and that of the optimal lower bound is 13.77 as compared to 148.4 for the total cost of online heuristic and the optimal lower bound. 

SCWNO can adjust three parameters, $V$, $\epsilon_q$ and $\epsilon_d$ according to the smart grid or IoT application requirements. In Table \ref{table:1}, the cost of proposed Lyapunov algorithm and the cost of optimal offline lower bound for different combinations of $\epsilon_d$, $\epsilon_q$ and $V$ are given. We can observe that for applications which demand stringent constraints on delay, small values of $V$, and $\epsilon_d$ should be used. On the other hand, if data quality is more important then larger values of $\epsilon_q$ are more desirable. In any situation, the algorithm can be effectively used to exploit any amount of QoS flexibility that might be available in order to minimize the HPC leasing cost of SCWNO.         

\begin{table*}[htb]
\centering
\caption{Cost of proposed online algorithm and optimal offline lower bound for different combinations of $\epsilon_d$, $\epsilon_q$ and $V$}
\includegraphics[width=0.9\textwidth,height=.3\textheight]{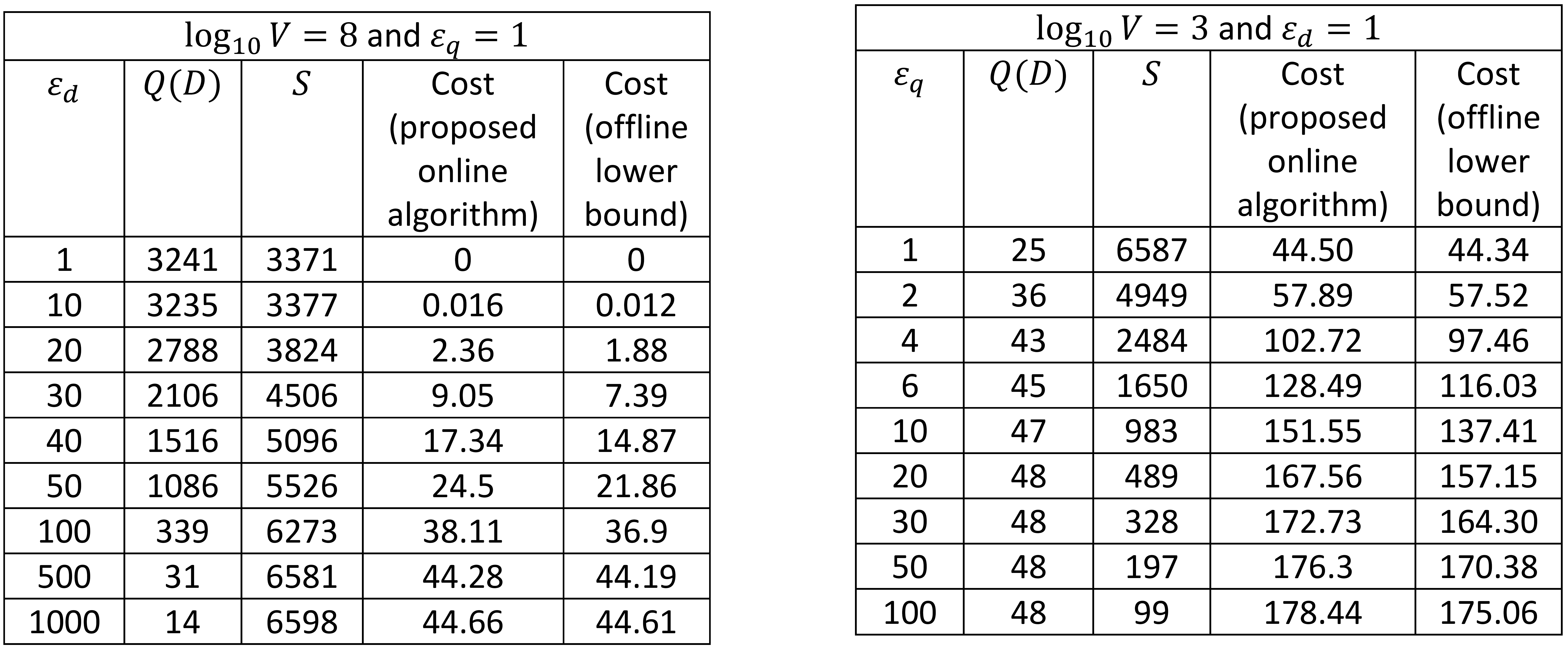}
\label{table:1}
\end{table*} 

\section{Conclusions}
\label{sec:conc}
In this paper, we considered the HPC leasing cost minimization problem of SCWNO operating in TVWS bands. We used the concept of delay and quality aware virtual queues and then used Lyapunov drift plus penalty function framework to develop an online algorithm. The algorithm can effectively tradeoff any amount of QoS flexibility that might be available in terms of queuing delays and data quality with the HPC leasing cost through three adjustable parameters. These parameters can be tuned according to the application requirements. Simulation results showed that the performance of the proposed algorithm is also very close to the optimal offline lower bound solution. This work can be further extended in future by relaxing the restrictions on the input arrival and departure rates from the queue and by allowing the simultaneous utilization of free and leased TVWS spectrum.

\appendix
\subsection{Derivation of per time slot optimization problems \eqref{master:eqn}}
We start with the one step Lyapunov drift function, i.e., 
\begin{equation*}
\begin{split}
\mathcal{L}(\Theta(t+1))-& \mathcal{L}(\Theta(t))= \frac{Q^2(t+1)-Q^2(t)}{2}+ \\ & \frac{Y^2(t+1)-Y^2(t)}{2}+\frac{Z^2(t+1)-Z^2(t)}{2} 
\end{split}
\end{equation*}
We can show that, 
\begin{equation*}
\begin{split}
\frac{Q^2(t+1)-Q^2(t)}{2} \leq & \frac{1}{2} [(R_{\text{max}}+A_{\text{max}})^2+A^2_{\text{max}}] + \\ &
 Q(t) [A(t)-R(t)]
\end{split}
\end{equation*}
\begin{equation*}
\begin{split}
& \frac{Y^2(t+1)-Y^2(t)}{2} \leq  \frac{1}{2} \max \left((\epsilon_q-1)^2, A_{\text{max}} \right) + \\ &
Y(t)\left[(\epsilon_q-1)(d_f^r(t)+d_p^r(t))-(d_f^f(t)+d_p^f(t)) \right]
\end{split}
\end{equation*}
\begin{equation*}
\begin{split}
\frac{Z^2(t+1)-Z^2(t)}{2} \leq & \frac{1}{2} \left(\epsilon_d-(\epsilon_d+1)R(t)\right)^2 + \\ &
Z(t)\left[\epsilon_d-(\epsilon_d+1)R(t) \right]
\end{split}
\end{equation*}
We can now write the conditional one step Lyapunov drift function as,
\begin{equation}
\begin{split}
\Delta & (\Theta(t)) \leq   B + Q(t) \mathbb{E} [A(t)-R(t) | \Theta(t)] + \\ &
Z(t)\mathbb{E} \left[\epsilon_d-(\epsilon_d+1)R(t) | \Theta(t) \right] + \\&
Y(t) \mathbb{E} \left[(\epsilon_q-1)(d_f^r(t)+d_p^r(t))-(d_f^f(t)+d_p^f(t)) | \Theta(t) \right]
\end{split}
\label{lupu:bound}
\end{equation}
where, $B=2.5+\frac{(\epsilon_d+1)^2}{2}+\frac{1}{2} \max \left((\epsilon_q-1)^2, 1 \right)$.
Adding, $V \mathbb{E} \left[d_p^f(t) c^f(t) + d_p^r(t) c^r(t) | \Theta(t) \right]$ to \eqref{lupu:bound}, we obtain,
\begin{equation}
\begin{split}
\Delta & (\Theta(t)) + V \mathbb{E} \left[d_p^f(t) c^f(t) + d_p^r(t) c^r(t) | \Theta(t) \right] \leq \\ &
B + V \mathbb{E} \left[d_p^f(t) c^f(t) + d_p^r(t) c^r(t) | \Theta(t) \right] + \\ &
Q(t) \mathbb{E} [A(t)-R(t) | \Theta(t)] + \\ &
Z(t)\mathbb{E} \left[\epsilon_d-(\epsilon_d+1)R(t) | \Theta(t) \right] + \\&
Y(t) \mathbb{E} \left[(\epsilon_q-1)(d_f^r(t)+d_p^r(t))-(d_f^f(t)+d_p^f(t)) | \Theta(t) \right]
\end{split}
\label{fin:lup}
\end{equation}
In the RHS of the above expression considering only the terms that involve the decision variables gives us $\mathcal{G}(t)$. As we can only control the decision variables, the problem is equivalent to minimizing $\mathcal{G}(t)$ subject to constraints on the optimization variables. 
%
%
\subsection{Proofs of Worst Case Bounds}
\textbf{Bound on $Q(t)$ \eqref{act:Q_b}:} We first prove that $Q(t)<V c^f_{\text{max}} +2, \forall t$. Since, $Q(0)=0$, it is true for $t=0$. We follow the approach adopted in \cite{lup_1,lup_2}. Suppose, it holds for time slot $t$. We show it also holds for time slot $t+1$. Consider the case when $Q(t) < V c^f_{\text{max}}$. Then $Q(t+1)<V c^f_{\text{max}} +2$, because the queue can only increase by 1 data unit at most by any time slot. Thus, the result holds in this case. Now consider the opposite case, when $V c^f_{\text{max}} < Q(t) < V c^f_{\text{max}} +2$. In this case, the value of $g_3(t)$ according to \eqref{val:3} will be negative. Therefore, the option of no transmission cannot be used and the algorithm will always transmit at least one data unit from the queue. There are three possible situations depending on the value of $h(t)$. In the best case, if $h(t)=2$, the algorithm will always transmit one full size data unit using the free TVWS spectrum since $Q(t) >0$. On the other hand, if $h(t)=1$, the value of $g_3(t)<0$ will result in the purchase of HPC spectrum for the transmission of full size data unit, if $g_2(t) > g_3(t)$, otherwise, the algorithm will exploit the free TVWS spectrum to transmit a reduced size data unit. Finally, in the worst case, if $h(t)=0$, once again as $g_3(t) <0$, the algorithm will purchase HPC spectrum for the transmission of one full size data unit if $g_3(t) < g_4(t)$, otherwise, the spectrum will be purchased for the transmission of one reduced size data unit. As we serve at least one data unit and the input arrival in the next time slot cannot exceed one data unit so the queue cannot exceed in the next time slot and $Q(t+1) \leq Q(t) <V c^f_{\text{max}} +2$. Finally, to complete the proof, if $Q(t)-1 \leq 0$, then from the queue dynamics, $Q(t+1)=A(t)\leq 1$, which is again less than $V c^f_{\text{max}} +2$. Therefore, $Q(t)<V c^f_{\text{max}} +2=Q_{\text{max}}, \forall t$.
\newline
\newline
\textbf{Bound on $Z(t)$ \eqref{act:Z_b}:} This proof is similar to the that of $Q(t)$. We first prove that $Z(t) < \frac{V c^f_{\text{max}}}{1+\epsilon_d} + \epsilon_d, \forall t$. Since, $Z(0)=0$, it is true for $t=0$. Suppose, it holds for time slot $t$. We show it also holds for time slot $t+1$. Consider the case when $Z(t) < \frac{V c^f_{\text{max}}}{1+\epsilon_d}$. Then $Z(t+1)<\frac{V c^f_{\text{max}}}{1+\epsilon_d} + \epsilon_d$, because this virtual queue can only increase by $\epsilon_d$ at most by any time slot (if the decisions is no transmission from non-empty queue). Thus, the result holds in this case. Now consider the opposite case, when $\frac{V c^f_{\text{max}}}{1+\epsilon_d} < Z(t) < \frac{V c^f_{\text{max}}}{1+\epsilon_d} + \epsilon_d$. In this case, because $Z(t)(1+\epsilon_d) > V c^f_{\text{max}}$, the value of $g_3(t)$ according to \eqref{val:3} will be negative. Therefore, the option of no transmission cannot be exercised and the algorithm will always transmit at least one data unit from the queue (even in the worst case when $h(t)=0$). As we must serve at least one data unit, the virtual queue $Z(t)$ will not increment (it only increments by $\epsilon_d$ in case of no transmissions from the non-empty queue) and hence in the next time slot $Z(t+1) \leq Z(t) < \frac{V c^f_{\text{max}}}{1+\epsilon_d} + \epsilon_d$. Therefore, $Z(t) < \frac{V c^f_{\text{max}}}{1+\epsilon_d} + \epsilon_d=Z_{\text{max}}, \forall t$. This completes the proof for $Z(t)$.
\newline
\newline
\textbf{Bound on $Y(t)$ \eqref{act:Y_b}:} Proof of this bound involves two interesting cases depending on the value of $\epsilon_q$. Following similar lines as in the proofs of the previous two bounds, we first show that $Y(t)<V c^f_{\text{max}} + \epsilon_q, \forall t$. Since, $Y(0)=0$, it is true for $t=0$. Suppose, it holds for time slot $t$. We show it also holds for time slot $t+1$. Consider the case when $Y(t) < V c^f_{\text{max}}$. Then $Y(t+1)<V c^f_{\text{max}} + \epsilon_q$, because this virtual queue can only increase by $\epsilon_q$ at most by any time slot. Thus, the result holds in this case. Now consider the opposite case, when $V c^f_{\text{max}} < Y(t) < V c^f_{\text{max}} +\epsilon_q$. In this case, the value of $g_3(t)$ according to \eqref{val:3} will be negative. Now, there are two cases depending on the value of $\epsilon_q$: i) $\epsilon_q > 1$ and ii) $0 < \epsilon_q \leq 1$. In the first case, $\epsilon_q Y(t) > Y(t)$, and therefore, $g_3(t) < g_2(t) < g_4(t)$. In this case, we will not transmit a reduced size data unit and in the worst case of $h(t)=0$, as $g_3(t) < 0$, we will purchase one full size data unit. Since we do not reduce data quality, the virtual queue $Y(t)$ will not increment (it only increments by $\epsilon_q$ when we transmit a reduced size data unit) and hence in the next time slot $(Y(t+1) < V c^f_{\text{max}} + \epsilon_q$. This completes the proof for the first case. In the second case, it is obvious that $g_2(t) < g_3(t)$ and $g_2(t) < 0$. Therefore, in this case, we can transmit a reduced size data unit. The transmission of a reduced size data unit will further increase the virtual queue length by $\epsilon_q$. However, at the same time, the departure of a reduced size data unit will also decrease the queue length by 1. Since, $\epsilon_q \leq 1$, the overall impact of departure and increment will either decrease the queue length or keep it constant. Thus, once again we will have $Y(t+1) < V c^f_{\text{max}} + \epsilon_q = Y_{\text{max}}, \forall t$. This completes the proof for the bound on $Y(t)$.  
\newline
\newline
\textbf{Bound on Queuing Delay \eqref{del:bound}:} In our model, at most one data unit can arrive or depart the queue in any time slot $t$. Due to this restriction, the worst case queuing delay is defined by the maximum backlog, i.e., $Q_{\text{max}}$. Therefore, any data unit that arrives in the queue at time slot $t$ will be out of the queue before the time slot $t+D_{\text{max}}+1$, where, $D_{\text{max}}=\left\lceil Q_{\text{max}}-1\right\rceil$, which leads us to,
\[D_{\text{max}}=\left\lceil V c^f_{\text{max}} +1 \right\rceil \]
and concludes the proof.
\newline
\newline
\textbf{Bound on Data Quality \eqref{qualy:bound}:} Let $S_{\text{max}}$ denote the number of reduced size data units that are transmitted in $D_{\text{max}}$ time slots. For all $\tau \in \{t+1,\ldots, t+D_{\text{max}} \}$, we have,
\[Y(\tau+1) \geq Y(\tau) - R(\tau) + \epsilon_q \left[d_f^r(\tau)+d_p^r(\tau)\right]  \] 
where, $d_f^r(\tau)+d_p^r(\tau)$ indicate the transmission of a reduced size data unit. We sum the above equation over $\tau \in \{t+1,\ldots, t+D_{\text{max}} \}$, which give us,
\[Y(t+D_{\text{max}}+1)-Y(t) \geq -\sum_{\tau=t+1}^{t+D_{\text{max}}} R(\tau)+ \epsilon_q \sum_{\tau=t+1}^{t+D_{\text{max}}} \left[d_f^r(\tau)+d_p^r(\tau)\right] \]
Using the maximum value of $R(\tau)=1, \: \forall \tau$ and substituting $\sum_{\tau=t+1}^{t+D_{\text{max}}} \left[d_f^r(\tau)+d_p^r(\tau)\right] = S_{\text{max}}$ in the above expression, 
\[Y(t+D_{\text{max}}+1)-Y(t) \geq -D_{\text{max}} + \epsilon_q S_{\text{max}} \]
Rearranging the terms and using the fact that $Y(t+1) \geq 0$ and $Y(t+D_{\text{max}}+1) \leq Y_{\text{max}}$, we can write,
\[S_{\text{max}} \leq \frac{Y_{\text{max}}+D_{\text{max}}}{\epsilon_q} \]
Finally, as the number of reduced size data units in $D_{\text{max}}$ time slots cannot exceed $D_{\text{max}}$ and further to ensure integer values we obtain,
\[S_{\text{max}} = \left\lceil \min \left(\left(\frac{Y_{\text{max}}+D_{\text{max}}}{\epsilon_q} \right),D_{\text{max}}\right) \right\rceil \]
which concludes the proof.

\bibliographystyle{ieeetran}   
\bibliography{references} 

\end{document}